\documentclass[sigconf]{acmart}

\usepackage{balance}
\usepackage{booktabs} % For formal tables
\usepackage{caption}
\usepackage{color}
\usepackage{fancyhdr}
\def\BibTeX{{\rm B\kern-.05em{\sc i\kern-.025em b}\kern-.08emT\kern-.1667em\lower.7ex\hbox{E}\kern-.125emX}}
\DeclareMathOperator*{\argmax}{arg\,max}

\copyrightyear{2019} 
\acmYear{2019} 
\setcopyright{acmlicensed}
\acmConference[]{}{}
\acmBooktitle{}

\begin{document}

\title{Stuck? No worries!: Task-aware Command Recommendation and Proactive Help for Analysts}

\author{Aadhavan M. Nambhi}
\authornote{These authors contributed equally to this research. The work was done when all the authors were affiliated with Adobe Research. Corresponding authors: \texttt{\{gaverma, burhanud\}@adobe.com} and \texttt{hs3673@nyu.edu}\\ 
~\\ \textcopyright  2019 Copyright held by authors. This is the authors' version of the work. A version of this work appeared in the Proceedings of \textit{27th Conference on User Modeling, Adaptation and Personalization (UMAP'19), June 9--12, 2019, Larnaca, Cyprus}.\\ \url{https://doi.org/10.1145/3320435.3320477} }
% \email{aadhavanm@cse.iitb.ac.in}
\affiliation{%
  \institution{IIT Bombay}
  \city{Mumbai, India}
}
\author{Bhanu Prakash Reddy}
\authornotemark[1]
% \email{bhanu@iitkgp.ac.in}
\affiliation{%
  \institution{IIT Kharagpur}
  \city{Kharagpur, India}
}

\author{Aarsh Prakash Agarwal}
\authornotemark[1]
% \email{aarshp@iitk.ac.in}
\affiliation{%
  \institution{IIT Kanpur}
  \city{Kanpur, India}
}

% \vspace{-1.0mm}
\author{Gaurav Verma}
% \email{gaverma@adobe.com}
\affiliation{%
  \institution{Adobe Research}
  \city{Bangalore, India}
}

\author{Harvineet Singh}
% \email{hs3673@nyu.edu}
\affiliation{%
  \institution{New York University}
  \city{New York, USA}
}

\author{Iftikhar Ahamath Burhanuddin}
% \email{burhanud@adobe.com}
\affiliation{%
  \institution{Adobe Research}
  \city{Bangalore, India}
}

%
% The abstract is a short summary of the work to be presented in the article.
\begin{abstract}
Data analytics software applications have become an integral part of the decision-making process of analysts. Users of such a software face challenges due to insufficient product and domain knowledge, and find themselves in need of help. To alleviate this, we propose a \textit{task-aware} command recommendation system, to guide the user on what commands could be executed next. We rely on topic modeling techniques to incorporate information about user's task into our models. We also present a help prediction model to detect if a user is in need of help, in which case the system \textit{proactively} provides the aforementioned command recommendations. We leverage the log data of a web-based analytics software to quantify the superior performance of our neural  models, in comparison to competitive baselines.
\end{abstract}

% \vspace{-5.0mm}


%
% A "teaser" image appears between the author and affiliation information and the body 
% of the document, and typically spans the page. 
\maketitle
% \vspace{-2.0mm}
\section{Introduction}
Powered by sophisticated computational techniques and powerful software tools, data analytics has seen a tremendous advancement in recent times. Users interact with data analytics software, such as \href{https://www.tableau.com/}{Tableau} and \href{https://powerbi.microsoft.com/}{Power BI}, to dissect and visualize data, and integrate results into their decision-making process. In several applications, data analytics software has become an essential tool that liberates users from tedious data processing tasks and allows them to focus on issues demanding more sophisticated human intelligence \cite{kandel2012enterprise}. 
% \HS{
However, interacting with such software is not an easy task, and novice analysts often find themselves lost \cite{blandford}. While querying data to create reports or building machine learning models, such as for user segmentation in the domain of website behavior analysis, analysts often face software-related problems which are further amplified by lack of support and in-person training \cite{mckinsey}. From the perspective of novice analysts, the workflows that are involved in such analytics applications are often complex sequence of commands and keeping track of them is difficult.
% 
% \Ifti{
We model the interaction of a user with an analytics interface in terms of \textit{commands} and \textit{tasks}. \textit{Commands} are the lowest level of interactions that a user can have with the UI (e.g., clicking on a button that sorts the data as per given column's values, drag-and-drop actions, etc.). We assume that the commands are executed in a sequence to achieve intermediate goals, called \textit{tasks}.
% T

\begin{figure}[!t]
\centering
    \includegraphics[width=1.0\columnwidth]{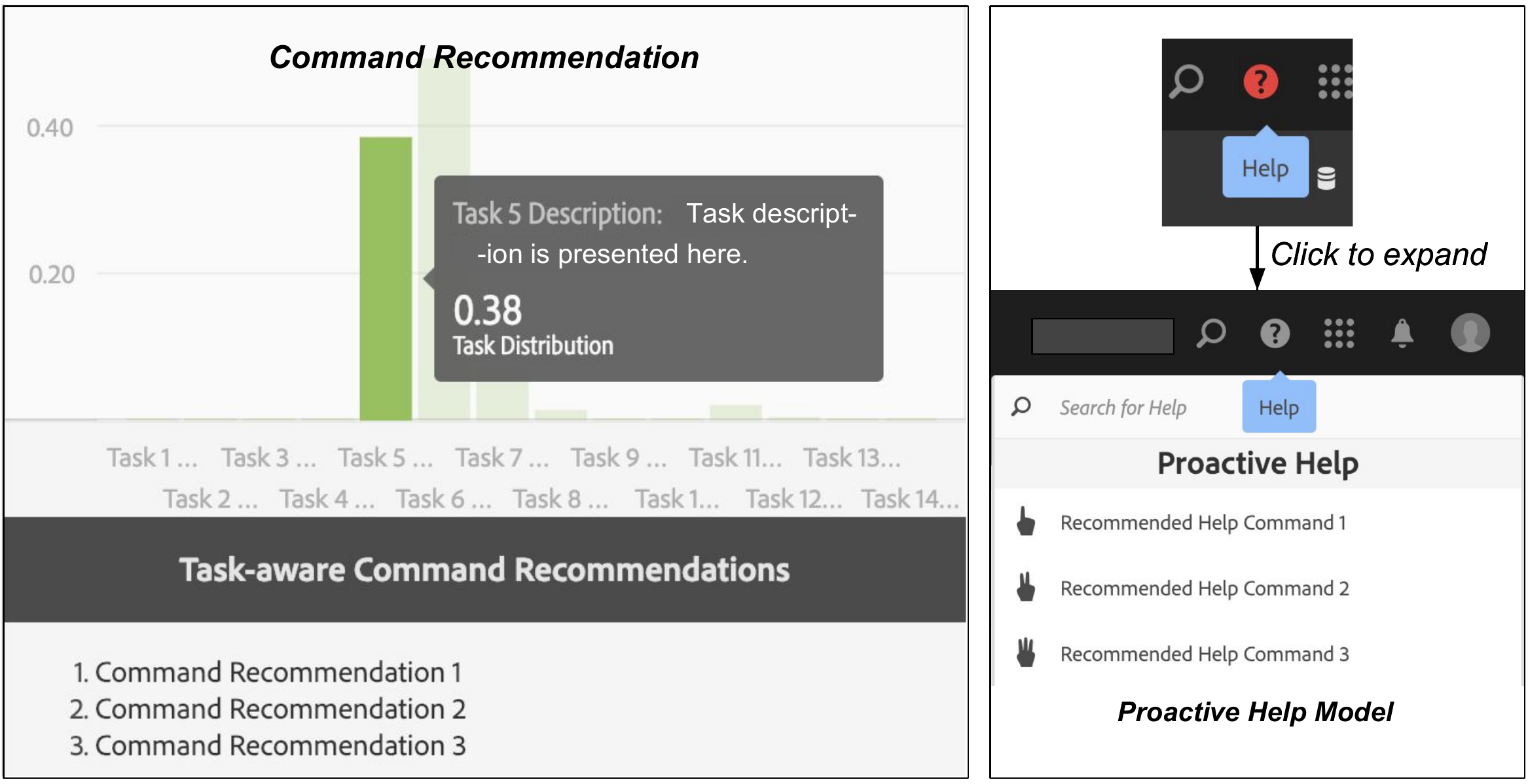}
    % \caption{}~\label{fig:figure1a}
    \vspace{-3.0mm}
  \caption{Left: The task-distribution of the ongoing sequence of commands is used to recommend future commands.
  Right: Proactive help is provided to the user.}
  \label{fig:representativeFigure}
  \vspace{-6.3mm}
\end{figure}

% \Ifti{
While significant research has been carried out to understand and analyze patterns in user behavior from application log data \cite{FrequentTasks, Workflows, LogAnalysis, davison1998predicting}, little research has been done to explicitly incorporate contextual information to recommend future commands \cite{React,RecommendWorkflows,CAPA, chen2018predictive}. By context, we mean commands leading up to the current activity. Additionally, limited amount of work has been done to proactively help the user in interacting with the interface. To this end, we present a novel approach that incorporates ongoing task information to recommend future commands, and provides proactive help to the user to alleviate their cognitive load while interacting with the interface.

% \Ifti{

Our work is inspired by \citet{RecommendWorkflows} who propose systems to recommend tutorial videos for complex software workflows. Their hierarchical approach operates at both task and command level. At the task level they consider topics found using a topic model as tasks, and at the command level they extract frequent command patterns for videos of each task by itemset mining. They build on a study of frequent user tasks from product log data \cite{FrequentTasks} and resort to topic modeling as a first layer to capture diverse usage patterns across the software. This is done to faithfully represent command patterns from less frequent tasks.

% \Ifti{

% \GV{

Following the prior art, we also use topic modeling techniques to model tasks that users can carry out by executing a sequence of commands. To minimize the data sparsity problem that topic modeling techniques like Latent Dirichlet Allocation \cite{LDA} introduce, we use Biterm Topic Modeling \cite{BTM}, which is designed to work on short texts.
% \HS{
Having inferred the tasks, we use this task information to recommend future commands for a given sequence (see Figure \ref{fig:representativeFigure}, right). 
The idea of incorporating current task distribution for recommending the next command is motivated by recent advancements in language modeling. In particular, our hypothesis is that combining sequence information along with the task distribution, which is analogous to coupling syntax and semantics in a language model, leads to better model performance \cite{TDLM, TopicRNN}.
%{

As we have mentioned earlier, interactions with computer interfaces almost always involve users executing a complex series of commands. To enhance users' experience, it would be useful if an intelligent interface could anticipate when the user is stuck, and provide help for the user to continue doing their tasks (see Figure \ref{fig:representativeFigure}, left). To this end, we propose a proactive help model.  Based on prior heuristics \cite{Lumiere}, there are some user activities, like inefficient command sequences, frequent searches, abrupt long pauses, frequently using undo commands, etc., that indicate a user's need of help. However, instead of explicitly modeling these heuristics as assumptions using a complicated rule-based model, we model them implicitly using data-driven approaches.

% W

% \
Our main contributions are three-fold: \textit{(i)} we propose a method to generate contextual command recommendations by incorporating ongoing task information, \textit{(ii)} we propose an LSTM-based method to detect if a user is in need of help, and \textit{(iii)} we comprehensively evaluate our proposed models to establish their superiority over competitive baselines. We believe our proposed methods will improve the quality of user interaction with data analytics software. 
% W
We discuss the dataset, proposed models, experiments, and evaluations in detail in the upcoming sections.

\vspace{-4.0mm}
\section{Dataset and Pre-processing}
\label{sec:Dataset}
The software application under consideration is a web-based analytics system, used to track, report, and analyze web traffic.
% T
We analyzed a proprietary dataset obtained from the usage logs of the interactive analysis and visualization section of the software. After pre-processing one month of usage logs from mid-May to mid-June 2018, we extracted around $350,000$ command sequences, for several tens of thousands of users.
% \H
Some of the command sequences with very little user activity 
% \H
were dropped. There were around 300 unique commands, which were obtained after dropping commands that were logged to indicate UI events and not explicitly executed by the users. 

We split the training and test sets on the basis of users, with roughly twenty percent of users in the latter set. During the pre-processing, to remove instances where the same command was executed several times consecutively, we limited the number of such consecutive occurrence of commands to two. The average number of commands per sequence, after pre-processing, was found to be 21. For uniformity, sequences with length less than 21 were dropped, and sequences with length greater than 21 were trimmed. Each of these sequences (henceforth, represented as $S$) are used as input to our command recommendation models.

%\I

For our proactive help models we use the same usage log data, but additional pre-processing steps were carried out beyond the steps discussed above. Out of the 300 commands, 14 commands were manually identified to denote users' requirement of help (for e.g., click on help icon). All sequences where any of these 14 commands occur for the first time, after $k$-th position in the sequence, were called as \textit{help sequences}. This particular step was taken to ensure that the first $k$ commands of such sequences can be used by our models to obtain some context before help is sought. Following this, the sequences that were identified as help sequences were trimmed to remove the portion following the help action, and then used as positive examples to train and test our models. In total, we had around $4,000$ positive examples. Non-availability of examples that could be explicitly called negative (i.e., help was not required), led us to using $20,000$ randomly sampled sequences that were not identified as help sequences. However, we acknowledge the downsides of this negative sampling --- there may be sequences in which users sought help using out-of-product search engines, or by other means. 
% \vspace{-3.0mm}
\section{Models}
% \vspace{-1.0mm}
% I
To recommend commands based on the current task information, we propose variants of sequence-to-sequence models \cite{seq2seq, karpathy2015char} which are compared against more traditional approaches that are focused around Markov models \cite{markov} and probabilistic suffix trees \cite{Workflows}. However, before recommending commands, the first part of the problem is to infer the ongoing task. We treat each of the command sequences, which are obtained from the usage log data, as a document and train a topic model with Biterm Topic Modeling (BTM). We use BTM, instead of more popular approaches for topic modeling such as Latent Dirichlet Allocation (LDA), to alleviate the data sparsity problem. This problem arises due to the co-occurrence matrix for each pair of commands being sparse as the command sequences are short in length. Following this, we use the obtained task distribution, denoted by $\mathbf{T}_S$ for command sequence $S$, to guide our command recommendation models.

% \I
% \vspace{-3.0mm}
\subsection{Command Recommendation Models}
% \H
We begin by describing conventional and competitive baseline approaches, and then move on to describing more sophisticated deep learning-based approaches.\\
\noindent\textbf{First-order Markov Model (FirstMM)}: A first-order Markov model computes the probability of the next action being $\mathbf{c}^i \in \mathcal{C}$, given the previous action $\mathbf{c}_t$, to make predictions about the next command $\mathbf{c}_{t+1}$. Mathematically,
% \vspace{-1.0mm}
% \[
	$\mathbf{c}_{t+1} = \argmax_{\mathbf{c}^i \in \mathcal{C}} \text{  } \Pr{(\mathbf{c}_{t+1} = \mathbf{c}^i \vert \mathbf{c}_{t})}$.
% \]
% \H
Here, $\mathbf{c}^i$ is a command belonging to set of all commands $\mathcal{C}$. More sophisticated models have been developed to enhance the performance of first-order Markov models.\\
\noindent\textbf{Variable length Markov Model (PST)}: Variable length Markov models consider variable number of previous commands as context instead of having fixed length contexts. Probabilistic Suffix Tree \cite{Workflows} is such a variable length Markov model. In a Probabilistic Suffix Tree (PST), the root node is assigned a `\texttt{null}' and every other node represents the sequence of commands that have to be executed in order to reach that node. The edge from a node to its children represents the probability of executing the next command in the sequence. Thus, given a sequence of commands which is represented as a node in a PST, we find the most probable future command by traversing the edge with highest probability from the node to its children.\\
% \B
\noindent\textbf{Task-aware Probabilistic Suffix Trees (TaskPST)}:
In order to make the predictions of a PST dependent on the task, we introduce the notion of Task-aware PSTs. For each task identified by BTM, we train one dedicated PST using sequences that are most likely to belong to that particular task. If there are $K$ tasks identified using BTM, this results in $K$ PSTs. At test time, a sequence is passed to all of the PSTs, and output from individual PSTs, which are probability distributions over the entire command vocabulary, are first weighted according to the task distribution of the test sequence, and then added to get the final output. This ensures that the final output, which is again a probability distribution over the entire command vocabulary, is influenced proportionately by the output of individual PSTs based on the task distribution of the sequence.

% \I
\noindent\textbf{Vanilla RNN (vRNN)}: With advancements in deep learning, several recurrent neural network (RNN) based approaches have been proposed to model sequential data, particularly natural language. 
% A 
One such class of approaches is referred to as sequence-to-sequence (seq2seq) modeling. These techniques have been successfully applied to tasks ranging from machine translation \cite{seq2seq} to query suggestion \cite{query}. Motivated by their wide-ranging applicability, we use variants of seq2seq models to recommend future commands. In our setting, a sequence of commands can be thought of as a sentence comprising words.
%T

%\B

Our proposed models use multi-layered Long Short-Term Memory (LSTM) \cite{LSTM} cells to encode the input sequence of commands into vectors of fixed dimensionality. These vectors are used by another LSTM (a decoder) to generate commands that align with the context of the sequence exposed so far. During the training phase, the generated command is compared to the ground truth command and the loss is backpropagated to update model parameters. Mathematically, at each unfolding of the decoder LSTM, the following probabilities are computed to generate the next command $\mathbf{c}_{t+1}$ in the sequence $S$:
% \I
% \vspace{-4.5mm}
\begin{equation}
\label{eq:lstm1}
	\Pr(\mathbf{c}_{t+1} = \mathbf{c}^i \vert \mathbf{c}_0, \mathbf{c}_1, \dots, \mathbf{c}_t)
% \vspace{-2.0mm}
\end{equation}
% \vspace{-0.5mm}
%\text{  } \forall i \in \{1, 2, \dots, \vert C \vert\}
Here, $\{\mathbf{c}_0, \cdots, \mathbf{c}_t$\} are the commands which are model's input at different timesteps, and $\mathbf{c}^i$ represents $i$-th command in the universal command set $\mathcal{C}$. This conventional seq2seq setup has been henceforth denoted as vRNN -- shortened form for vanilla RNN. 
%  \I

\noindent\textbf{Task-aware RNN (TaskRNN)}:
Our first modification to vRNN involves concatenating the task distribution $\mathbf{T}_{S}$ for a sequence $S$, obtained using BTM, with the trainable vector embeddings of commands, given by $\mathbf{c}_j$.
% \H
Consequently, Equation \ref{eq:lstm1} is modified to: $\Pr(\mathbf{c}_{t+1} = \mathbf{c}^i \vert \mathbf{x}_0, \dots, \mathbf{x}_t)$, where $\mathbf{x}_j = \mathbf{c}_j \oplus \mathbf{T}_{S}$. The notation $\oplus$ denotes vector concatenation.
%\text{  } \forall i \in \{1, 2, \dots, \vert C \vert\}
Serving ongoing task information as an additional input provides a broader context to the model and also helps in filtering out commands that may not be related to the current task. It is worth noting that while we have access to the entire sequence of commands at training time to determine the task distribution, it is unreasonable to assume the same at testing time. Therefore, during the testing phase, we only use the task distribution of the sequence seen so far by our model, as opposed to using the task distribution of the entire sequence as done in the training phase. Due to this heterogeneity of model inputs in the training and testing phase, we were motivated to explore variants that do not assume access to task distribution of the entire sequence. 

\noindent\textbf{Joint Task and Command RNN (JTC-RNN)}:
% \H
This model comprises of two sub-modules. The first module is given as input a new command $\mathbf{c}_{t}$ in the sequence at timestep $t$, along with the task distribution of the sequence seen so far $S^t$, denoted by $\mathbf{T}_{S^t}$. This module is responsible for predicting the task distribution if the whole sequence $S$ were seen, which is written as $\mathbf{\hat{T}}_{S}^t$. During the training phase, the predicted task distribution is compared with the ground truth task distribution $\mathbf{T}_{S}$ using Kullback-Leibler divergence \cite{KL}. Furthermore, at each timestep, the output of this module, i.e., $\mathbf{\hat{T}}_{S}^t$, is concatenated with the trainable vector embeddings of current command in the sequence $\mathbf{c}_t$ and is used by the second module to predict next command in the sequence $\mathbf{c}_{t+1}$. In the light of this modification, Equation 1 can now be re-written as: $
	\Pr(\mathbf{c}_{t+1} = \mathbf{c}^i \vert \mathbf{x}_0', \dots, \mathbf{x}_t') \text{  }$
%\forall i \in \{1, 2, \dots, \vert C \vert\}
, where $\mathbf{x}_j' = \mathbf{c}_j \oplus \mathbf{\hat{T}}_{S}^j$. Since $\mathbf{\hat{T}}_{S}^j$ is computed using the part of the command sequence seen so far $S^j$, there is no heterogeneity between the training and test phase. 

% \I

% \I
% \vspace{-3.5mm}
\subsection{Proactive Help Models}
Our proposed help model dynamically monitors a user's interaction with the interface and, if need be, proactively recommends them to seek help along with providing command recommendations. This proactive help recommendation is provided by turning the help icon to red, as shown in Figure \ref{fig:representativeFigure}. 

We formulated proactive help recommendation as a supervised binary classification problem, for which the data comprises of positive (i.e., help sequences) and negative sequences as discussed in Section \ref{sec:Dataset}. We train a Random Forest classifier and an LSTM classifier for this classification problem. 
For each of these models, to incorporate the temporal information, we experiment with concatenating the time interval $\Delta_t$ between execution of current command $\mathbf{c}_t$ and the previous command $\mathbf{c}_{t-1}$, and the trainable vector embedding of the current command $\mathbf{c}_t$. During training, the LSTM classifier takes the concatenation of time interval $\Delta_j$ and $\mathbf{c}_j$ as input at each timestep $j$, and the output of the \textit{final} unfolding of LSTM is passed through a fully-connected neural (FCN) layer to output binary class probabilities, given by: $
	\Pr(y_t = class \text{ } \vert \text{ } \mathbf{c}_0 \oplus \Delta_0, \dots, \mathbf{c}_t \oplus \Delta_t) \text{,}
$
% \vspace{-1.0mm}
where $class \in \{help, no\_{help}\}$. During test time, all the settings are the same as that in the training time, except that the output of \textit{every} unfolding (after capturing sufficient context using first $k$ commands of the sequence) is passed through FCN layer to output the class probabilities. Incorporating time intervals between consecutive commands along with the command embeddings, allows the model to implicitly learn some of the heuristics that we mentioned earlier. We briefly discuss these observations in Section \ref{sec:evaluation}.
% \vspace{-3.0mm}
\section{Experiments}
% \I

For inferring tasks using BTM, standard Gibbs sampling \cite{Gibbs} is used to compute the values of the involved multinomial distributions. The two Dirichlet priors $\alpha$ and $\beta$ are hyperparameters, and a low value for them will give rise to sharper distributions. We choose $\alpha$ and $\beta$ to be 0.001 and 0.005 respectively. Another user-defined parameter for topic models is $K$, which is the number of tasks. For the software under consideration, based on our experiments, we had the most coherent tasks for $K = 14$. Following task identification, each sequence $S$ had a $14$-dimensional task distribution associated with it, denoted by $\mathbf{T}_{S}$. For PST and TaskPST
% \
, we limit the maximum depth of every PST to $10$. To limit branching of the trees, we only consider sequences that appear more than a specified number of times in the training data; we chose the threshold to be $7$. 
%O

For our proposed RNN-based models
% \
, we consider a sequence $S$ to be made up of commands $\{\mathbf{c}_0, \dots, \mathbf{c}_N\}$, where $N$ is chosen to be 20, as discussed in Section \ref{sec:Dataset}. Each command $\mathbf{c}_i$ is represented as a $32$-dimensional embedding which can be trained along with the rest of the model to give a semantically rich representation of the command. At each time step $T = t$, this command embedding $\mathbf{c}_t$ is either input to a multi-layered LSTM-encoder directly (vRNN), or concatenated with either $\mathbf{T}_{S}$ (TaskRNN) or $\mathbf{\hat{T}}_{S}^t$ (JTC-RNN) and then input to the multi-layered LSTM encoder to predict the next command in the sequence, i.e., $\mathbf{c}_{t+1}$. During training, we minimize cross-entropy loss $\mathcal{L}_1$
%b 
However, since JTC-RNN has an additional sub-module that estimates task-distribution at every timestep, it has an additional loss component -- apart from $\mathcal{L}_1$ -- which is given by KL divergence of estimated task distribution $\mathbf{\hat{T}}_{S}^t$ with respect to the ground truth task distribution $\mathbf{T}_{S}$.
% i

Both of our proactive help models operate on sequences that have a minimum context length (i.e., $k$) of 8. 
As mentioned previously, this has been done to ensure that the models have sufficient context before help is sought.
For Random Forest classifiers (which comprise of $50$ decision trees), to reduce the dimensionality of input vectors, we project an identity matrix of size $\vert \mathcal{C} \vert \times \vert \mathcal{C} \vert$ using Gaussian random projection \cite{GaussianRP} to a matrix of size $\vert \mathcal{C} \vert \times 8 $. Here $\vert \mathcal{C} \vert$ is the cardinality of the command set and the value $8$ is a design choice. This gives us an 8-dimensional representation of each unique command in the vocabulary, which is concatenated with corresponding time interval in seconds. The LSTM classifier again uses 8-dimensional trainable vector embeddings concatenated with the time interval. The binary cross-entropy loss is backpropagated to update the model parameters. 
%W
Across our all models, we use Adam optimizer \cite{adam}, with a learning rate initialized at $10^{-3}$. Also, we train our models with early stopping based on accuracy over validation set. While presenting the results in Table 1 and 2           , we provide the average of quantified values over $5$ different runs. 
%T
\vspace{-2.0mm}
\section{Evaluation}
\label{sec:evaluation}
To evaluate the quality and coherence of identified tasks by BTM, which was trained in an unsupervised fashion, we rely on the assessment of two experts who have several years of experience with the software under consideration. These experts were shown top 20 commands for each of the 14 tasks. They were able to identify and relate an actual task that an analyst does while interacting with the software interface for almost all tasks. Eight task labels matched across both the experts, which included comparing data across time periods, creating clusters, dashboard building and analysis, and visualizing data.

% \I
To quantify the performance of command recommendation models, we use Top-1 and Top-5 accuracy. We rank order and compare the top 1 (and top 5) recommended command(s) by these models with next command in ground truth sequences to evaluate these accuracies. 
For uniformity, the test set across all these models was exactly the same. 
Table 1 summarizes the results of command recommendation models. Note that the superior performance of the TaskRNN model can be attributed to utilizing the task distribution of the entire sequence when recommending commands. Given that such information about the task distribution will not be available during model deployment, the results of JTC-RNN can be deemed as realistic performance.\\

\vspace{-5.0mm}
\begin{table}[!h]
\centering
\caption{Performance of command recommendation models}
\vspace{-4.0mm}
\scalebox{0.88}{
\setlength\tabcolsep{3.0pt}
\begin{tabular}[!h]{ c|c|c|c|c|c|c}
\label{temp:accuracies}
 \textbf{Accuracy} & \textbf{FirstMM} & \textbf{PST} & \textbf{TaskPST} & \textbf{vRNN} & \textbf{TaskRNN} & \textbf{JTC-RNN}  \\  
 \hline
 \textbf{Top 1}& 0.443 & 0.563 & 0.565 & 0.538 & \textbf{0.620} & 0.575 \\ 
 \textbf{Top 5}& 0.697 & 0.659 & 0.665 & 0.766 & \textbf{0.897} & 0.792 \\\hline
\end{tabular}
}
\end{table}
% \
\vspace{-2.0mm}
As it can be observed in Table 1, models that incorporate task-information, in general, perform better than those that do not. Task information guides the process of recommending next command to make it more apropos.\\
% \vspace{-0.5mm}
Help models have been evaluated using precision and recall, and the trade-off between the two has been summarized using area under the receiver operating characteristic (AU-ROC) curve. \\

% \H
\vspace{-5.0mm}
\begin{table}[!h]
\centering
\caption{Performance of proactive help models}
\vspace{-3.0mm}
\scalebox{0.85}{
\setlength\tabcolsep{3.0pt}
\begin{tabular}{ c|c|c|c}
\label{tab:help_model}
 {\textbf{Help Prediction Models}} & {\textbf{Precision}} & {\textbf{Recall} }  & {\textbf{AU-ROC}} \\
 \hline
 Random Forest (Commands only) & 0.20 $\pm$ 0.03 & 0.28 $\pm$ 0.04 & 0.73 $\pm$ 0.08 \\ 
 Random Forest (Time $\oplus$ Commands) & 0.23 $\pm$ 0.02 & 0.30 $\pm$ 0.02  & 0.74 $\pm$ 0.04  \\ 
 LSTM Classifier (Commands only) & 0.26 $\pm$ 0.04 & 0.37 $\pm$ 0.03  & 0.81 $\pm$ 0.11 \\ 
LSTM Classifier (Time $\oplus$ Commands) & \textbf{0.27} $\pm$ 0.06 & \textbf{0.40} $\pm$ 0.05  & \textbf{0.83} $\pm$ 0.13  \\\hline 
\end{tabular}
}
\end{table}
% \H
\vspace{-2.0mm}
From Table 2, it can be observed that \textit{(i)} LSTM-based classifier performs better than Random Forest baseline, and \textit{(ii)} concatenating time interval $\Delta_t$ leads to an improvement in results for both the classifiers. A small value of standard deviation in Table 2 indicates consistent experimental runs. For qualitative analysis, we manually examined the sequences that were predicted as true positives by our classifiers. Here, we encountered instances where commands were repeated in loops, and sequences that contained frequent search commands. These observations align with existing heuristics developed to model help scenarios \cite{Lumiere}.
% F
% \vspace{-2.0mm}
\section{Conclusion and Future work}
In this paper, we investigated the effectiveness of incorporating task information while recommending future commands to the user. The results of our task-aware command recommendation models, when compared to conventional task-agnostic models, are quite promising and call for future explorations along this line of work. We also propose an LSTM-based method to detect if a user is in need of help.
% W
Our quantitative and qualitative evaluations suggest that the help model is capable of implicitly modeling some of the heuristics that have existed in the literature for quite some time. 

In future work, we will experiment with more sophisticated models that can incorporate the two lines of our work, i.e., task-aware command recommendation and proactive help, in a single end-to-end model, and evaluate their efficacy.
% \vspace{-1.5mm}
\section{Acknowledgements}
The authors would like to thank the anonymous reviewers, Sanjeev Biswas, Brandon George, Prakhar Gupta, Nate Purser,  Abhilasha Sancheti and Atanu Sinha for their inputs and comments on this work.

% W
\balance
\bibliographystyle{ACM-Reference-Format}
\bibliography{acmart}

% 
% If your work has an appendix, this is the place to put it.

\end{document}